# The disaster of the Nazi-power in science as reflected by some leading journals and scientists in physics

## A bibliometric study

*Manuel Cardona* and Werner Marx*
*Max Planck Institute for Solid State Research, Stuttgart (Germany)*

**Abstract**

The dramatic consequences of the Nazi-power for science are described extensively in various articles and books. Recent progress in information systems allows a more quantitative reflection. Literature databases ranging back to the beginning of the 20th century, the ISI citation indexes ranging now back to 1945 and sophisticated search systems are suitable tools for this purpose. In this study the overall break in the scientific productivity and that of selected physical journals are examined. An overview of the citation impact of some 50 leading physicists is given. The productivity before and after departure is analyzed and connected to biographical data.

**Introduction**

The indescribable agony of the victims of the Nazi-terror, in particular the Jews, exceeds any imagination and cannot be quantified. More comprehensible are the consequences for politics, economics, and science. The impact on the influence and reputation of German science and the shift of the centers of excellence and the language of science are well known. The loss of scientific potential and talent to Germany has been extensively discussed but mostly in a speculative way. In this study, a well established method for evaluating present science was used to analyze the past. We have determined the citation impact since 1945 of the publications of some of the most influential emigrants in physics. The break in productivity is measurable when relating the more recent impact to the publication years of the cited papers of the emigrants. In consideration of their hard fate and their huge contribution even to current science they should never fall into oblivion.

**Methods**

The data presented here are mostly based on the Science Citation Index (SCI) marketed by Thomson Scientific (the former ISI - Institute for Scientific Information). The findings result from the Web of Science (WoS), the search platform provided by ISI, as well as the database SciSearch (SCI under the host STN International). The search options available under STN make it possible to perform extensive citation analyses. Currently, SciSearch only covers the source items (citing papers) since 1974, whereas the WoS stretches back to 1945 now. However, the cited papers (references) therein stretch back before 1945. The data of table 1 and figures 7 and 8 were established by combining WoS and STN.

* Corresponding author:
Prof. Dr. Manuel Cardona
Max Planck Institute for Solid State Research
Heisenbergstrasse 1, D-70569 Stuttgart, Germany
E-mail address: m.cardona@fkf.mpg.de

**Overall scientific productivity**

Recently, the literature file of the Chemical Abstracts Service (CAS) of the American Chemical Society was extended back to 1907, the year when CAS was founded. This file covers not only chemical literature, but also a large amount of literature from the life sciences and physics as well. Therefore, the time curve of the publications per year covered by this file visualizes well the break in scientific productivity caused by World War I and II (see figure 1).

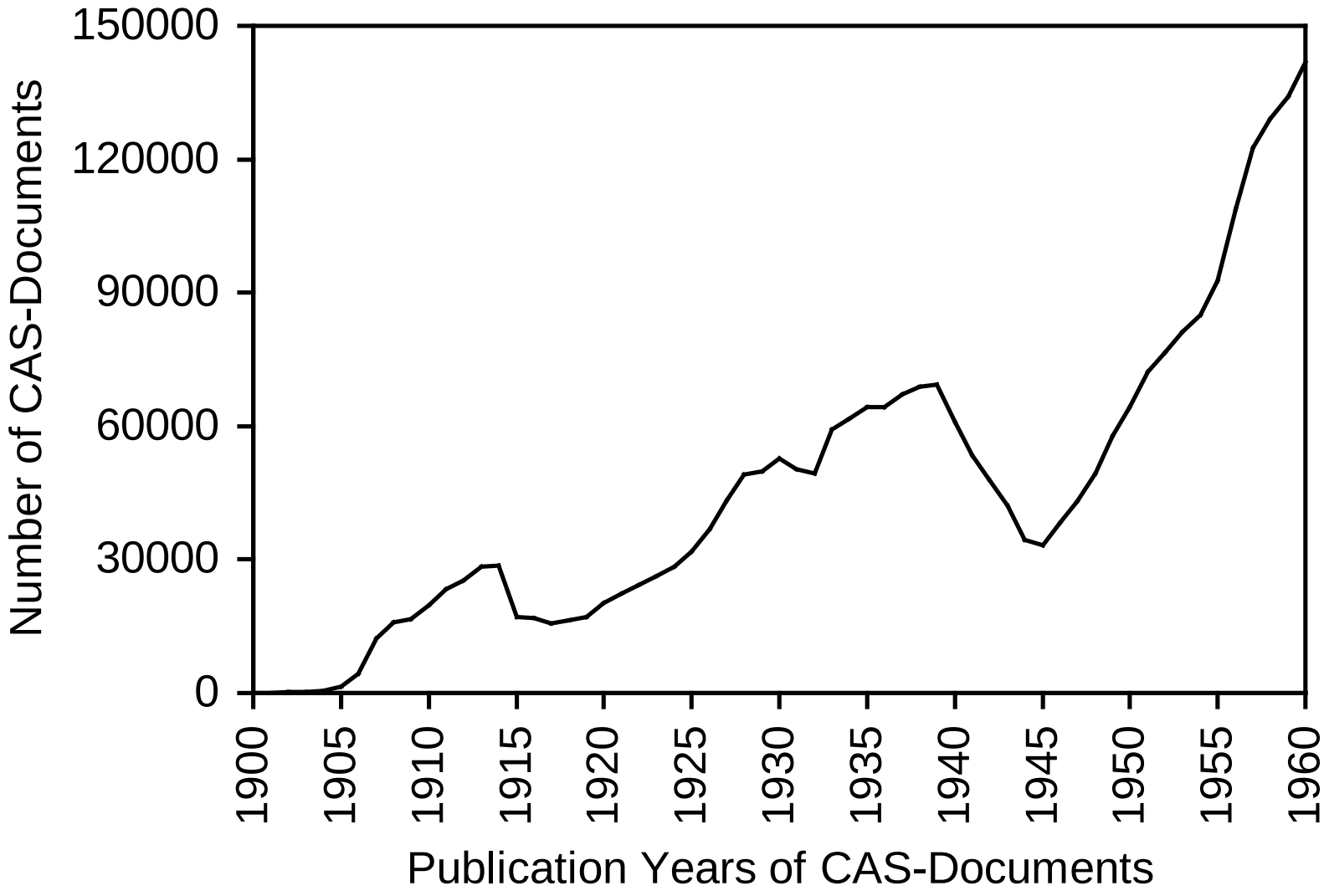

**Figure 1:** Distribution of publication years of the documents (patents included) published in the time period 1907-1960. From the literature file of the Chemical Abstracts Service (CAS) of the American Chemical Society.

The war-caused breaks are also reflected by the age distribution of the overall pre-1960 publications cited by more recent articles. The age distribution for the time period 1900-1960 of all references cited within all SCI documents since 1974 was determined (see figure 2). Compared with the publication time curve the reference age curve is smoothed out somewhat by the time gap between the date of publication and citation.

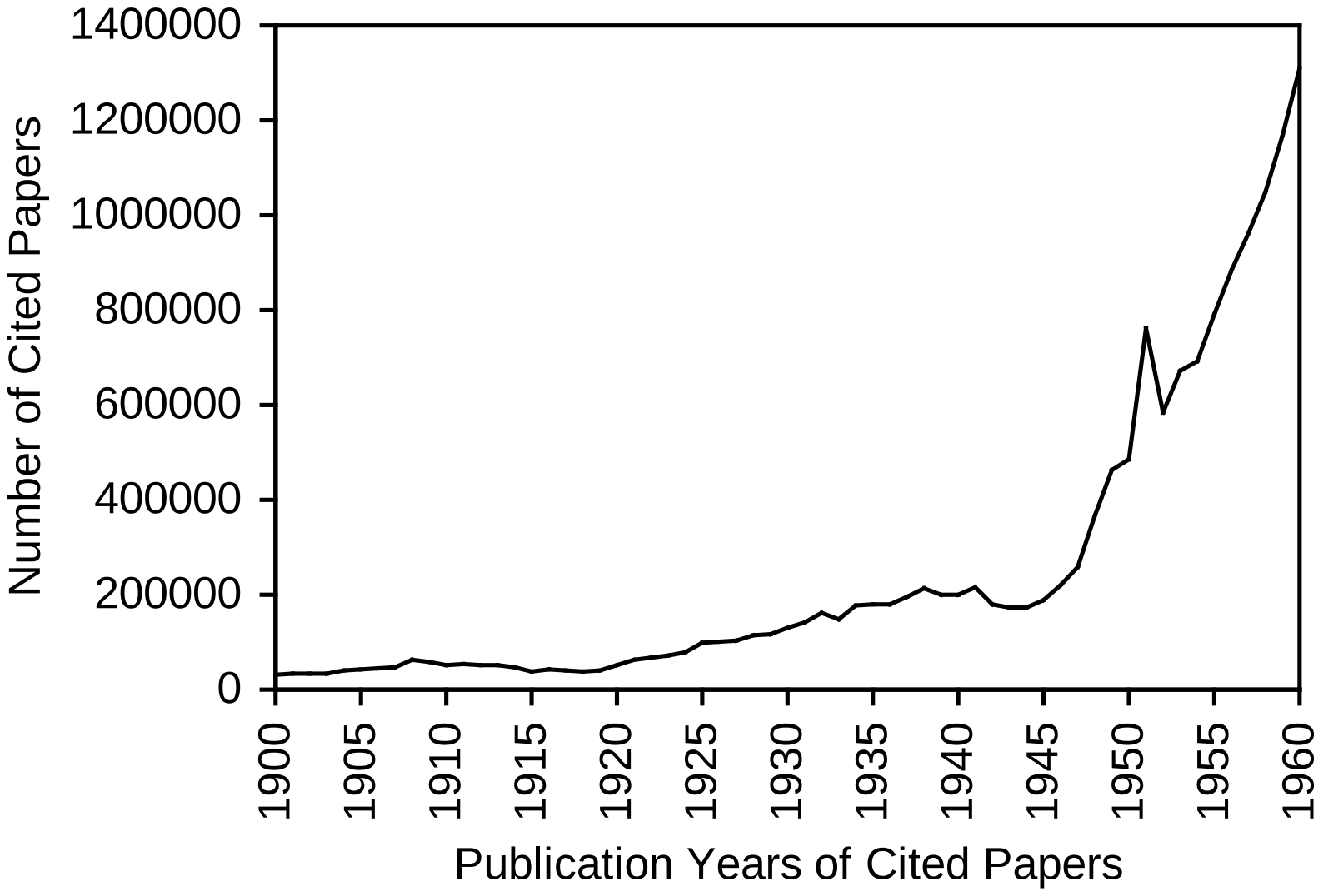

**Figure 2:** Age distribution of references published between 1900 and 1960 and cited in SCI documents published in the time period 1974-2004. The peak at publication year 1951 corresponds to the most-cited scientific paper of all times (O.H. Lowry: J. Biol. Chem. V193 P265 (1951), about 280,000 citations).

## Publication numbers of leading journals

As in the case of the overall scientific output, the CAS literature file can also be utilized to determine the time-dependent output of single journals. Although the articles of physics journals are not covered completely in the CAS file, no distortion of the time curve is expected. The time-dependent number of publications per year was determined for two leading German physics journals (Annalen der Physik, Zeitschrift für Physik) in the time period 1907-1970. The output of these journals decreased from some hundred publications per year to about zero in 1945 (see figures 3 and 4). In comparison with these journals the output decrease of Physical Review and other US journals is less dramatic.

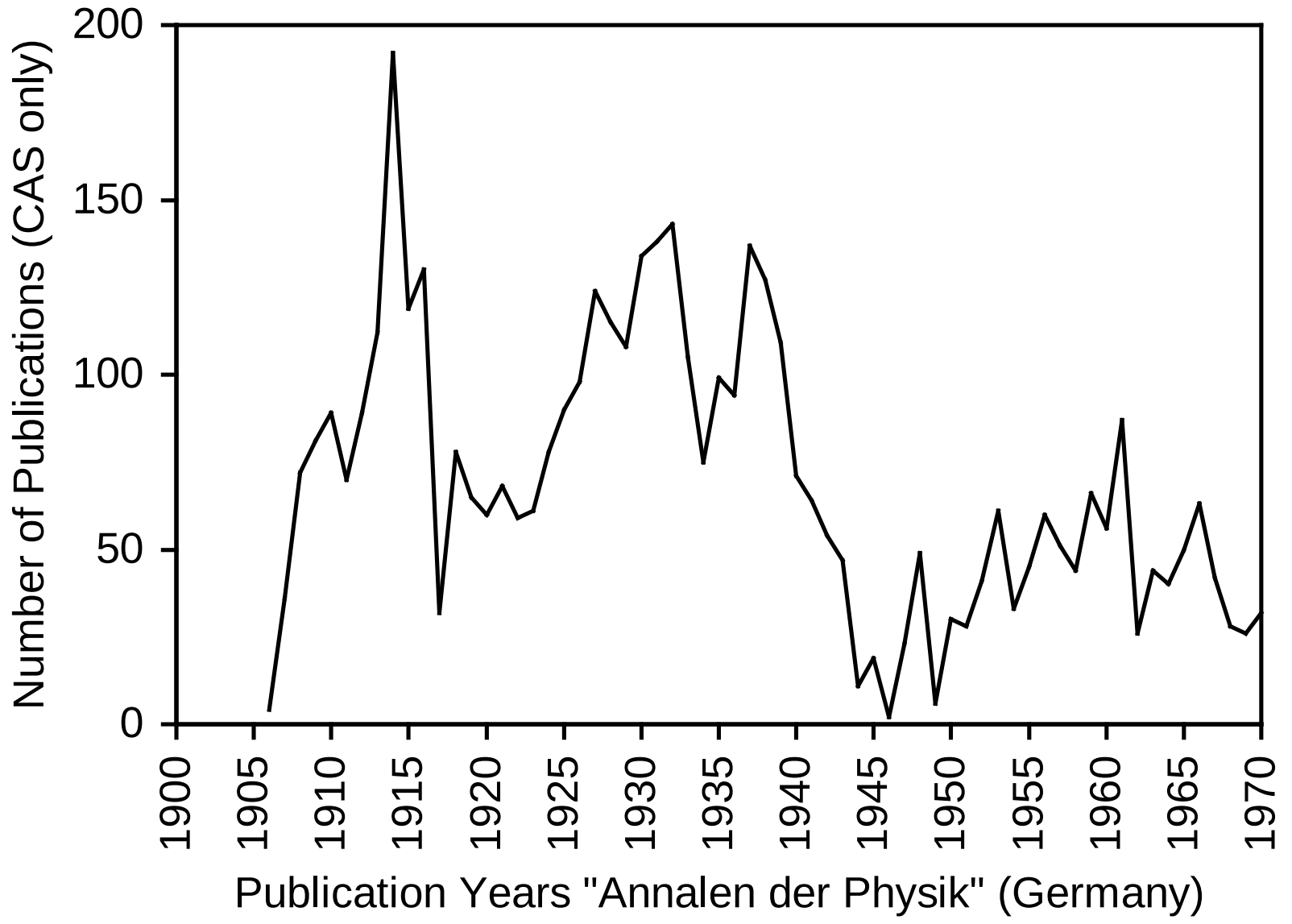

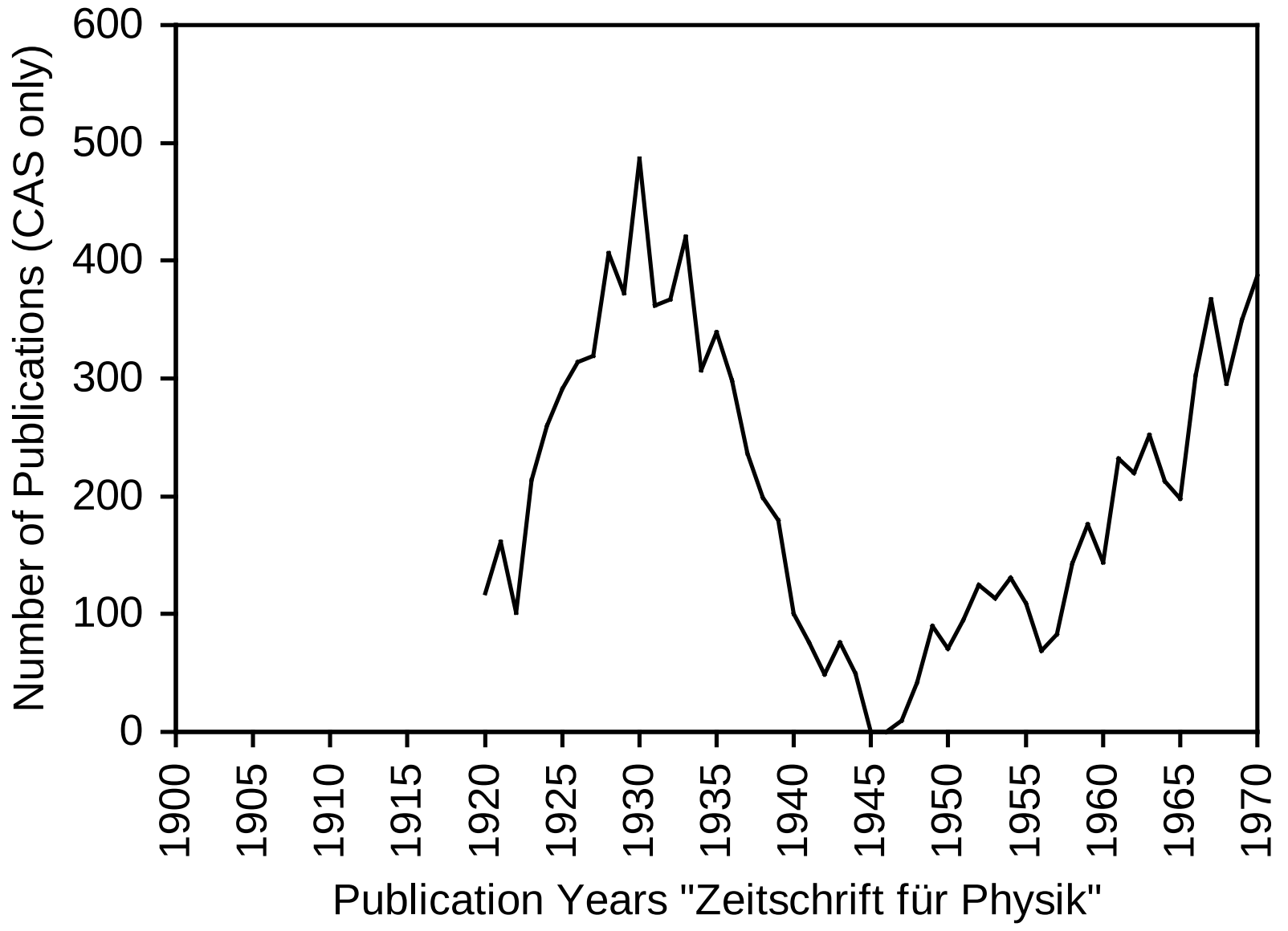

**Figures 3 and 4:** Time-dependent number of publications of the two leading German physics journals in the time period 1907-1970 (CAS documents only).

**Impact of early journals in recent articles**

The output decrease of a specific journal should be reflected by a similar decrease of its citation impact, although the citations occurred decades after publication. In order to test this conjecture, the reference age distribution of citations within SCI documents published since 1974 to pre-1970 articles of some leading physics journals was analyzed. The impact break is most dramatic for Zeitschrift für Physik which published many articles from the golden age of physics before World War II (see figure 5). As in figure 2 the number of citing papers (instead of the number of citations) was determined. Note that one citing paper may include more than one citation (reference) of articles from a specific journal and the same publication year (e.g. two articles from Zeitschrift für Physik both published in 1930).

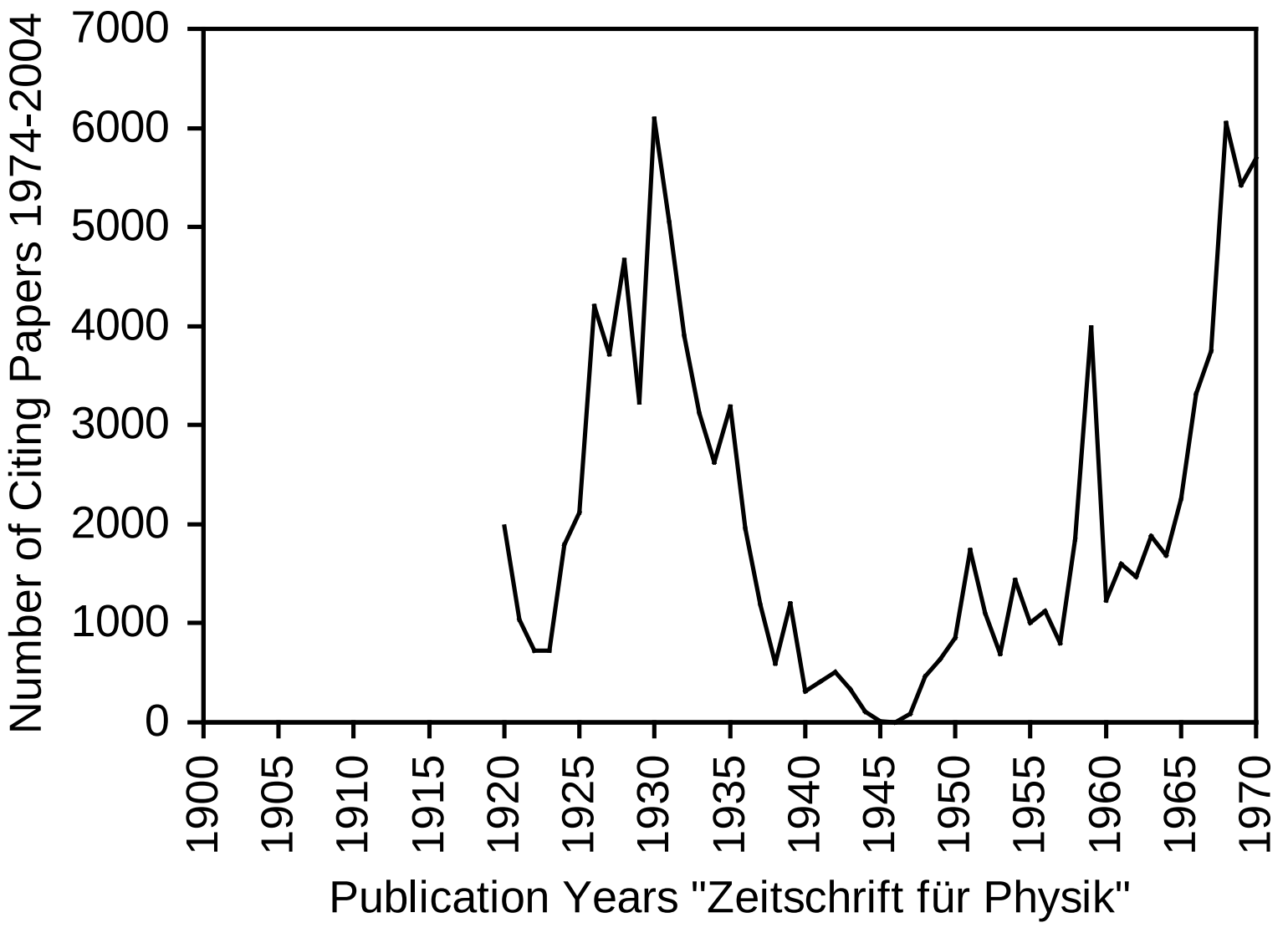

**Figure 5:** Time-dependent number of citing papers since 1974 of the publications of Zeitschrift für Physik published in the time period 1920-1970.

**Impact of some fifty influential emigrants**

There are many articles and books concerning scientists who had to leave Germany after the Nazis came to power (see references). Furthermore, the internet offers a large amount of bibliographic information. Fifty-two emigrants were selected for analyzing their citation impact in more detail (see table 1).

| **Emigrants** | **Age in the Year 1945** | **Citing Papers 1945-2004 all Publ.** | **Citing Papers 1945-2004 1900-1944 Publ.** |
|---|---|---|---|
| Heinz/Henry H.Barschall (1915-1997) | 30 | 1669 | 71 |
| Hans A. Bethe* (1906-) | 39 | 16262+ | 11787 |
| John M. Blatt (1921-1990) ns | 24 | 8182 | 5 |
| Felix Bloch* (1905-1983) | 40 | 7987 | 4572 |
| Max Born* (1882-1970) | 63 | 32185+ | 9025 |
| Wolfgang J. Choyke (1926-) | 19 | 4868 | 0 |
| Peter Debye* (1884-1966) | 61 | 15647 | 9734 |
| George V. De Hevesy* (1885-1966) | 60 | 2583 | 1393 |
| Martin Deutsch (1917-2002) ns | 28 | 5028 | 154 |
| Paul P. Ewald (1888-1985) | 57 | 3231 | 2960 |
| Albert Einstein* (1879-1955) | 66 | 15646 | 12397 |
| Kasimir Fajans (1887-1975) | 58 | 1242 | 991 |
| James Franck* (1882-1964) ns | 63 | 2482 | 1139 |
| Otto R. Frisch (1904-1979) | 41 | 467 | 145 |
| Herbert Fröhlich (1905-1991) | 40 | 8523 | 906 |
| Dennis Gabor* (1900-1979) | 45 | 4547 | 61 |
| Richard Gans (1880-1954) | 65 | 1173 | 998 |
| Maurice Goldhaber (1911-) | 34 | 6655 | 48 |
| Gertrude Goldhaber-Scharff (1911-1998) | 34 | 3404 | 19 |
| Walter Gordon (1893-1939) | - | 654 | 654 |
| Samuel A. Goudsmit (1902-1978) | 43 | 915 | 693 |
| Fritz Haber* (1868-1934) | - | 2869 | 2869 |
| Walter Heitler (1904-1981) | 41 | 6164 | 1743 |
| Gustav Hertz* (1887-1975) ns | 58 | 601 | 201 |
| Gerhard Herzberg* (1904-1999) | 41 | 27148+ | 36 |
| Viktor F. Hess* (1883-1964) ns | 62 | 450 | 203 |
| Walter Kohn* (1923-) | 22 | 24952 | 5 |
| Nicholas Kurti (1908-1998) | 37 | 529 | 187 |
| Rudolf Ladenburg (1882-1952) | 63 | 921 | 731 |
| Rolf Landauer (1927-1999) | 18 | 8467 | 4 |
| Fritz London (1900-1954) | 45 | 5514 | 3936 |
| Heinz London (1907-1970) | 38 | 1030 | 255 |
| Hermann F. Mark (1895-1992) ns | 50 | 7623 | 1158 |
| Liese Meitner (1878-1968) | 67 | 364 | 298 |
| Kurt Mendelssohn (1906-) | 39 | 1223 | 254 |
| Rudolf Minkowski (1895-1976) | 50 | 2352 | 424 |
| Lothar W. Nordheim (1899-1985) | 46 | 1830 | 979 |
| Abraham Pais (1918-2000) | 27 | 4500 | 27 |
| Wolfgang K.H. Panofsky (1919-) | 26 | 2910 | 20 |
| Wolfgang Pauli* (1900-1958) | 45 | 6272 | 3395 |
| Rudolf E. Peierls (1907-1995) | 38 | 8803 | 2753 |
| Erwin Schrödinger* (1887-1961) | 58 | 5951 | 4467 |
| Sir Francis E Simon (1893-1956) ns | 52 | 1954 | 967 |
| Hertha Sponer (1895-1968) | 55 | 1436 | 801 |
| Walter E. Spear (1921-) | 24 | 5249 | 0 |
| Frank Stern (1928-) | 17 | 6061 | 0 |
| Otto Stern* (1888-1969) | 57 | 2291 | 1892 |
| Harry Suhl (1922-) | 23 | 5047 | 1 |
| Leo Szilard (1898-1964) | 47 | 1932 | 458 |
| Edward Teller (1908-2003) | 37 | 12408 | 723 |
| Viktor F. Weisskopf (1908-2002) | 37 | 9330 | 2770 |
| Eugene P. Wigner* (1902-1995) | 43 | 18893 | 10030 |
| Total | | ca. 330,000 | ca. 100,000 |

*  Nobel laureates, ns = namesakes
\+  Citing papers 1974-2004 only, because of system limits
Date of searching: 24.06.04 - 14.07.04

**Table 1:** Physicists who departed positions in Germany during the time period 1933-1945. Some younger scientists who left Germany before finishing their formal education (J.M. Blatt, W.J. Choyke, R. Landauer, F. Stern, and H. Suhl) are included.

The number of citing papers (instead of the number of citations, one citing paper may include more than one citation) were established using the WoS. The citing papers cover the time period 1945-2004. Currently there is no citation index available reaching back before 1945. The cited papers (i.e. the references related to the papers of the analyzed scientists) alternatively cover all publication years (3rd column) and only the papers published in the time period 1900-1944 (4th column).

Some high-impact scientists (with more than about 5000 citations) exceed the WoS system limits. In that case the citing papers were estimated under WoS only for the period from 1945 till 1973 and then completed under STN for the period from 1974 till 2004. The citing papers of co-author publications (the selected author is not the first author) are only included if the papers are WoS source items (i.e. no pre-1945 co-author papers are covered).

The citation data established under STN for the citation time period 1974-2004 only include the citations of the first author publications. Thus, the given overall numbers of citing papers related to some high-impact authors are lower limits. However, most of the early papers have only one single author and no substantial loss of citations can be expected. The citing papers of three of the most cited scientists (Bethe, Born, and Herzberg) could only be determined under STN for the time period since 1974 because of the WoS system limits.

Some scientists have namesakes. An effort was made to purge their papers by using the cited reference tables of the WoS. If the data of authors with namesakes had to be completed under STN (Franck, Simon), reference publication years after the year of death were excluded. After their emigration, some scientists preferred using middle name initials (instead of simply the first names they used before 1945). The author searching was conducted with truncation after the first initial. By this procedure all versions were included.

Note that the given citation numbers are restricted to the citation time period since 1945 and therefore the impact of the earlier papers published before 1945 cannot be completely determined. The large overall impact of some scientists (Bethe, Born, Debye, Fröhlich, Heitler, Herzberg, Peierls, Wigner) is partially due to highly cited books. References to these books are included in our count provided they appeared in ISI source journals.

**Emigrants: total impact, single papers**

The overall impact of all 52 emigrants of table 1 has also been determined. The number of citing papers increased from 5000 in the year 1974 to about 8000 at present (see figure 6). As the physical disciplines increased by about the same amount, we may conclude that there is an unusually persistent interest involving the work of the emigrants being analyzed here. This exemplifies the great loss to their countries of origin, which probably still affects today their scientific standing.

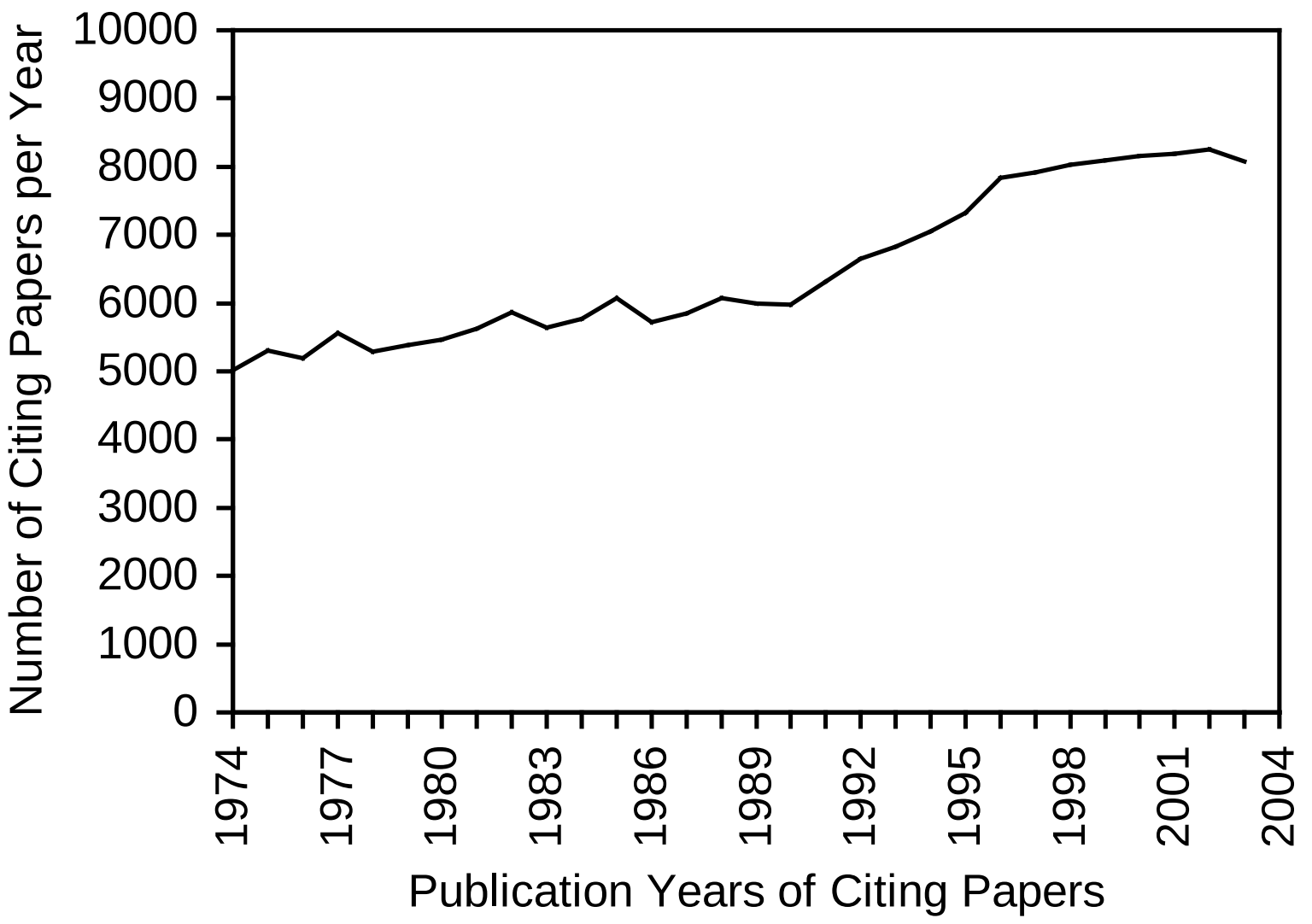

**Figure 6:** Time-dependent overall impact of the 52 emigrants of table 1.

The seminal papers of science are now so well known that they appear in textbooks. They are taken for granted and scientists mostly do no longer refer to the original articles. For example the original papers of Albert Einstein's theory of relativity are well cited now. However, his name and the theory of relativity are mentioned much more frequently. Thus, the overall impact of the emigrants is not completely documented by the citations of their papers and cannot be quantified exactly.

Some interesting information is obtained from the time-dependence of the citations of single articles. The time-dependent citation graph for an article is sometimes called its citation history and may be viewed as the sales figure of that article. The citations generally do not increase substantially until one year after publication. They normally reach a maximum after about three years and then usually decrease, accumulating citations at a lower level. However, some highly important emigrant articles show a distinctive time delay, which signifies that they were ahead of their time (see figures 7 and 8).

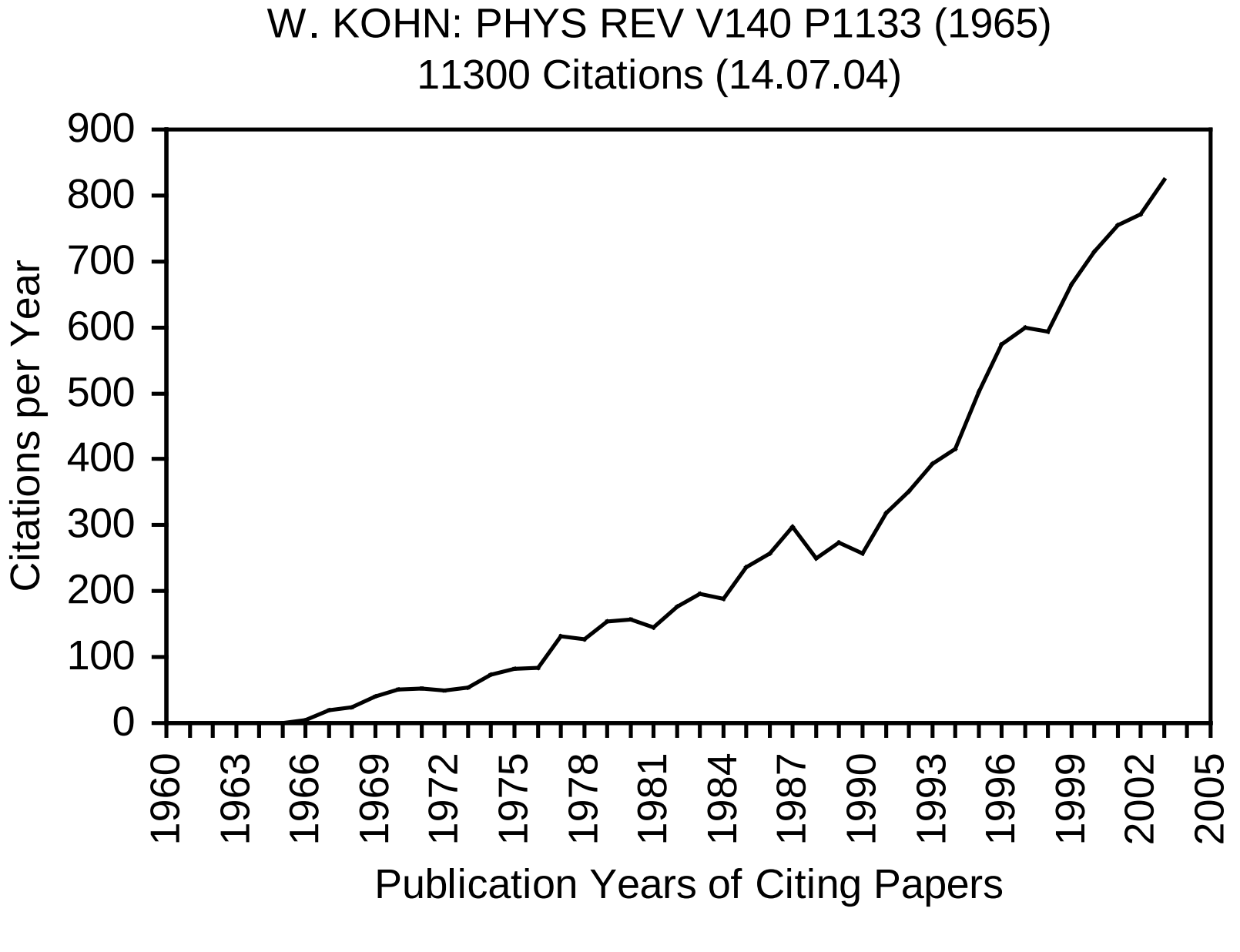

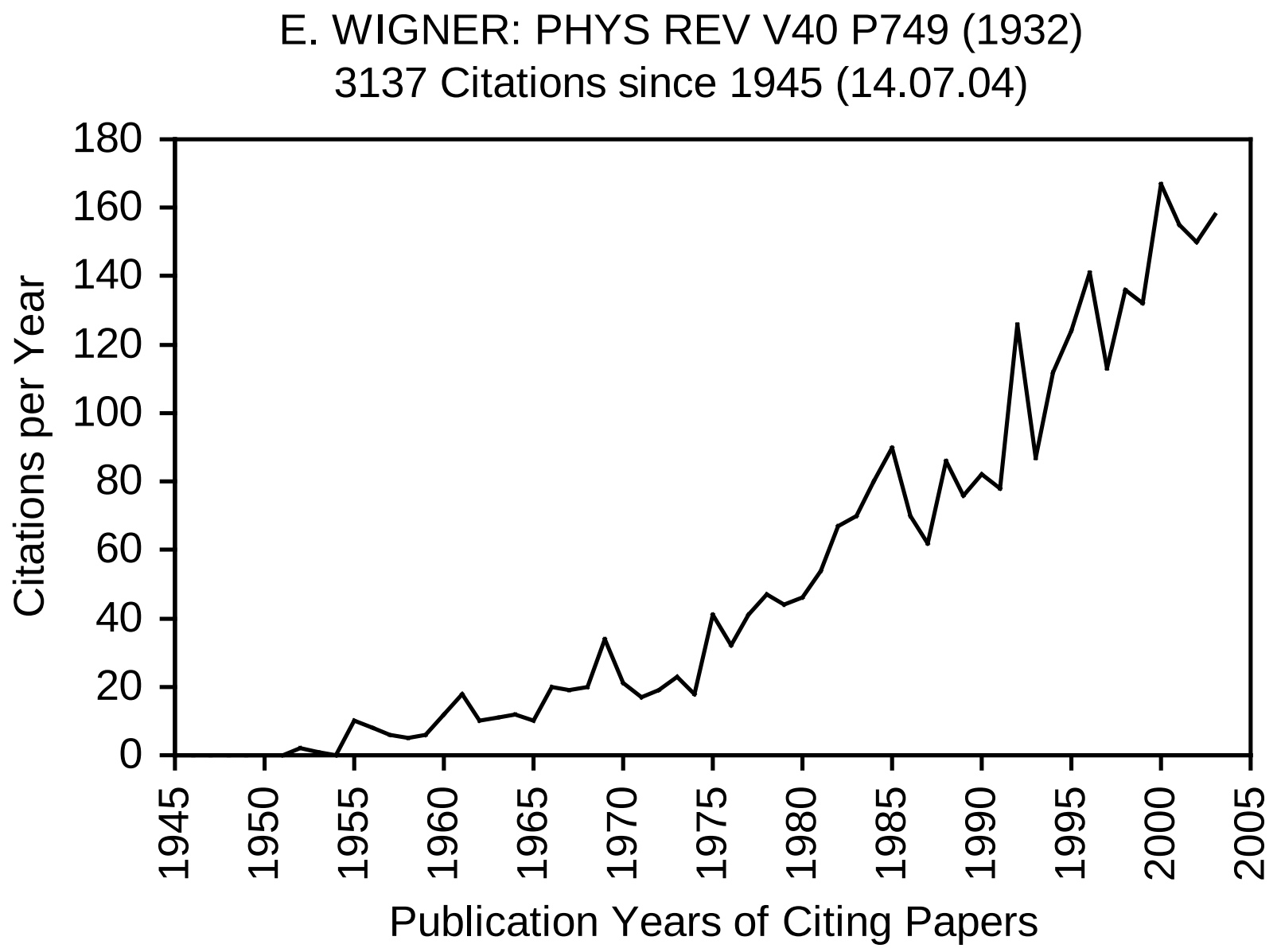

**Figures 7 and 8:** Citation histories of two highly-cited papers of Walter Kohn and Eugene Wigner, a young and a senior emigrant. The citations of both papers are highly delayed in time.

**Emigrants: paralyzed or successful careers**

Citation bar diagrams with the numbers of citations (since 1945) as function of the publication years of the cited papers (e.g. the papers of the emigrants) show the relative contribution of publication years (and papers) to the overall impact (see figures 9 and 10). When analyzing the emigrants of table 1 by this method, we find two distinct types of immigrants: those with a considerable scientific impact before emigration and those who left as young children or students, who made their careers largely in their new countries. The latter can be easily identified by the small number of citing papers in the right hand column of table 1 (Blatt, Choyke, Kohn, Landauer, Spear, F. Stern ... and maybe also Barschall, Goldhaber, Goldhaber-Scharff, Pais, and Panofsky). They also must be viewed as a considerable loss of talent to their countries of origin.

Among the more senior emigrants the majority of those on table 1 were able to continue a productive career, several of them an extremely productive one crowned in some cases by the Nobel Prize. A few of them did not do so well. This is revealed by comparing the last two columns in table 1. In some cases (Goudsmid) they moved to a highly successful career in science management. A few (Lise Meitner, Otto Frisch, Gustav Hertz, Otto Stern) seem to have had difficulties adapting to a new, sometimes hostile environment. Fritz Haber, a stauch Prussian soul, never recovered from the blow and succumbed shortly after emigrating.

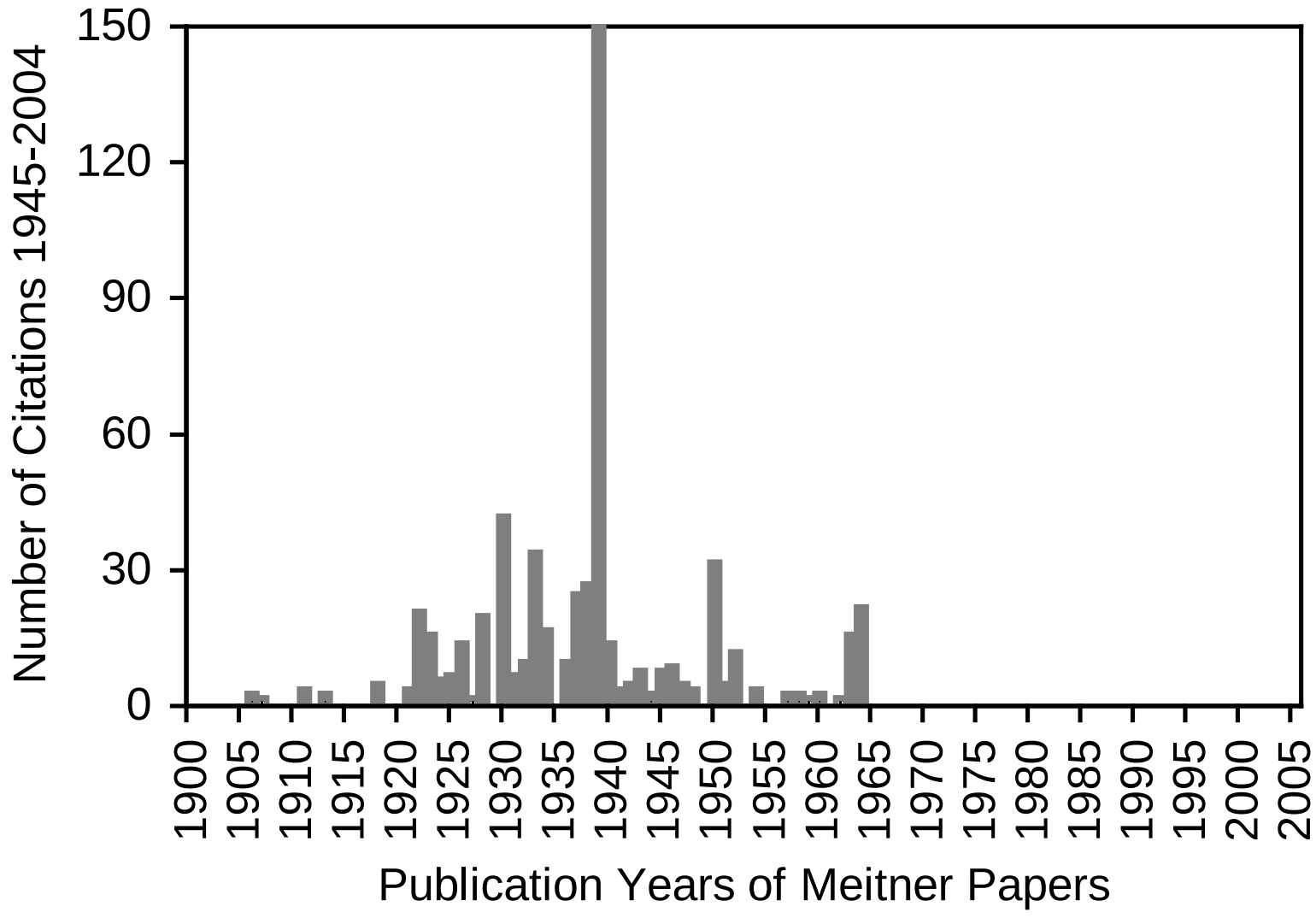

**Figure 9:** Citation bar diagram of Liese Meitner with the number of citations since 1945 as function of the publication years of the cited papers (i.e. the Meitner-papers).

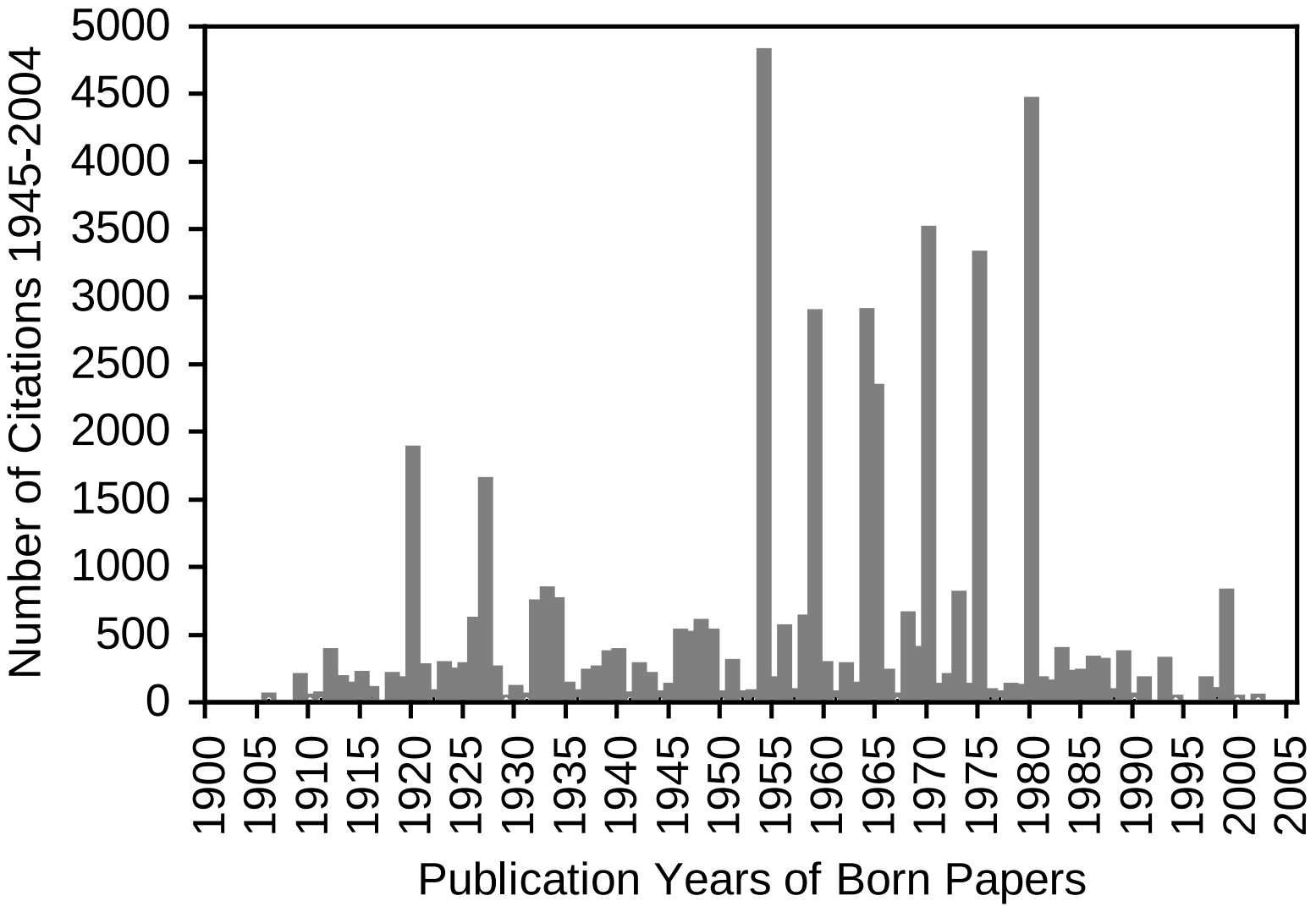

**Figure 10:** Citation bar diagram of Max Born with the number of citations since 1945 as function of the publication years of the cited papers (i.e. the Born-papers). The impact past 1970 (the year of his death) is mainly due to the various later editions of his book "Principles of Optics" (altogether about 23,000 citations).

## Conclusions

The break in the scientific productivity during the Nazi-period is visualized by publication and reference numbers. The citation impact data of selected emigrants in physics reflect the loss of talent to Germany and the other countries of origin. The large contribution of emigrants to modern physics is remembered. Such analysis enables a more quantitative view of the overall disturbance of science by the Nazi-power.